\documentstyle[12pt]{article}

\newcommand{\Rho}{{\rm P}}
\newcommand{\MIPT}{
                   \small
                   \it Moscow Institute of Physics and Technology,\\
                   \small
                   \it Institutsky per. 9, Dolgoprudny, Moscow region
                  }
\font\twmsbm=msbm10 scaled 1200
\font\nmsbm=msbm9
\newfam\bbold
\textfont\bbold=\twmsbm
\scriptfont\bbold=\nmsbm
\scriptscriptfont\bbold=\nmsbm

\font\twscr=rsfs10 scaled 1200
\font\nscr=rsfs10
\newfam\script
\textfont\script=\twscr
\scriptfont\script=\nscr
\scriptscriptfont\script=\nscr
\def\Cal#1{{\fam\script#1}}
\author{M.~G.~Ivanov\thanks{e-mail: mgi@landau.ac.ru} \\ \MIPT }
\date{October 31, 1997}
\title{\vskip-20mm
       \hfill {\small hep-th/9710110}
       \vskip20mm
       p-brane solutions
       and Beltrami-Laplace operator
      }
\begin{document}
\maketitle

\begin{abstract}
  Generalization of the harmonic superposition rule
  for the case of dependent choice of harmonic
  functions is given. Dependence of harmonic functions
  from all (relative and overall) transverse coordinates
  is considered using the Beltrami-Laplace operator.
  Supersymmetry of IIB 10D supergravity solutions
  with only non-vanished 5-form field and 11D
  supergravity solutions is discussed.
\end{abstract}

\section{Introduction}
 The p-brane solutions are the subject of numerous researches
 \cite{Stelle}--\cite{LuPope}.

 In the recent time the considerable progress in the
 understanding of harmonic superposition rule was
 achieved
 (for review see \cite{Stelle,OBZOR} and references therein).
 Two different approaches to the problem was
 elaborated. The first approach is initiated by the investigations
 of the supersymmetry properties of the solutions
 (see \cite{TS}--\cite{Ts212}).
 The second approach is initiated by the investigations
 of the Einstein field equations
 (see \cite{AV}--\cite{Angle}).

 The present paper continue the research, which was made in
 the papers \cite{AIR,AIV}.
 In these works the case
 of independent choice of harmonic functions was
 mainly considered.
 The condition
 \begin{equation}
 \label{plain}
   (\partial_{i_1}H)(\partial_{i_2}H)\not=0
   ~\Longrightarrow~
   g_{i_1i_1}=g_{i_2i_2}
 \end{equation}
 was necessary condition of dependence
 of harmonic function from coordinates $x_i$.

 The main subject of the present paper is consideration
 of the case of dependent choice of harmonic functions
 and elimination of restriction (\ref{plain}).

 The following metric represent the example of p-brane solution
without restriction (\ref{plain}) in the case of 11D supergravity
 \begin{eqnarray}
 \label{ex0-1}
   &&ds^2=(H_1H_2)^\frac{1}{3}
          \left(
           -(H_1H_2)^{-1}dx_0^2
%          \right.\\ \nonumber
           + H_1^{-1}(dx_1^2+dx_2^2)
           + H_2^{-1}(dx_3^2+dx_4^2)
           +
%            \left.
             \sum_{i=5}^{10} dx_i^2\right),\\
 \label{ex0-2}
         &&\Box H_1(x_3,\dots,x_{10})=0,~~~~~~~~~~
           \Box H_2(x_5,\dots,x_{10})=0,
 \end{eqnarray}
 where the symbol ``$\Box$'' designate
 the Beltrami-Laplace operator.
   The Beltrami-Laplace operator is a covariant object,
 the using of it let us possibility of more profound
 understanding of p-brane solutions properties.
   The system of equations (\ref{ex0-1}), (\ref{ex0-2})
 is nonlinear system.
   The simplest way to build the example of
 the function $H_1$ is to set
 \begin{eqnarray}
 \label{r1}
 H_1=H_{11}(x_3,x_4)H_{12}(x_5,\dots,x_{10}),\\
 \label{r3}
 \eta_{KL}\partial_K\partial_L H_{Ru}=0,~~~R=1,2~~~u=1,2.
 \end{eqnarray}

\section{General results}
 Let us consider the case of arbitrary number of
 antisymmetric and dilatonic fields in the space-time
 of arbitrary dimensions and signature. The Kaluza-Klein
 theory with extra time-like dimensions has been
 considered in \cite{AV85}.
 The appropriate action has the following form
 \begin{equation}\label{act}
   I=\frac{1}{2\kappa ^{2}}
      \int d^{D}X\sqrt{|g|}
      \left(
         R-\frac{1}{2}(\nabla\vec\phi)^2-
         \sum_{I=1}^{k}
         \frac{s_I e^{-\vec\alpha^{(I)}\vec\phi}}{2(d_I+1)!}
         F^{(I)2}
      \right),
 \end{equation}
 where $s_I=\pm 1$, $\vec\phi$ is a set of the dilaton fields,
 $F^{(I)}$ is a field of $d_I+1$-form.

 We can write extremal and non-extremal solutions
 in terms of incidence matrices by the following way.
 Let $\Delta_{aK}^{(I)}$ and $\Lambda_{bK}^{(I)}$
 are the electric and magnetic incidence matrices
 respectively, the elements of these matrices
 are zeros or unities. The matrices describe
 the antisymmetric fields:
 \begin{eqnarray}
 \label{f-sum}
	 F^{(I)}&=&\sum_{a=1}^{E_I}dA_a^{(I)}
          +\sum_{b=1}^{M_I} F_b^{(I)},\\
 \label{ea}
   A^{(I)}_{a~M_1\cdots M_{d_I}}&=&
     \epsilon_{M_1\cdots M_{d_I}}h_a  H_a^{-1}
     \prod_{i=1}^{d_I}\Delta_{aM_i}^{(I)},\\
 \label{ma}
   F_b^{(I)~M_0\cdots M_{d_I}}&=&
     \epsilon^{M_0\cdots M_{d_I} K}
     h_b|g|^{-1/2}e^{\vec\alpha^{(I)}\vec\phi}
     \partial_K  H_b^{-1}
     \prod_{i=0}^{d_I}\Lambda_{bM_i}^{(I)},
 \end{eqnarray}
 where $\epsilon^{01\cdots}=\epsilon_{01\cdots}=1$ are
 totally antisymmetric symbols.

 We shall use also the {\it general incidence matrices}:
 the {\it field matrix} $\Delta_{RK}$ and
 the {\it brane matrix} $\Upsilon_{RK}$,
 \begin{equation}
%  \footnotesize
   \Delta  =\left(
              \begin{array}{c}
                \Delta^{(1)}\\
                \vdots\\
                \Delta^{(k)}\\
                \Lambda^{(1)}\\
                \vdots\\
                \Lambda^{(k)}
              \end{array}
            \right),\qquad
   \Upsilon=\left(
              \begin{array}{c}
                \Delta^{(1)}\\
                \vdots\\
                \Delta^{(k)}\\
                1-\Lambda^{(1)}\\
                \vdots\\
                1-\Lambda^{(k)}
              \end{array}
            \right).
 \end{equation}
 Index $R$ numerate branes, and index $K$ numerate coordinates.

 Let
 \begin{eqnarray}
 \label{ds2}
    ds^2&=&\sum_{K,L=0}^{D-1}
           \left(
             \prod_{R}
               H_R^{\varsigma_R h_R^2
                    \left\{
                       \Delta_{RK}-\frac{d_R}{D-2}
                    \right\}
                   }
           \right)
            \eta_{KL}dX^K dX^L,\\
 \label{fi}
   \vec\phi&=&1/2\sum_{R} \varsigma_R \vec\alpha_R h_R^2 \ln H_R.
 \end{eqnarray}
 In all above equations $h_R$ are constants,
 and $H_R$ are functions from coordinates.
 $\varsigma_R=\pm 1$: $\varsigma_R=+1$ for magnetic branes,
 $\varsigma_R=-1$ for electric branes,
 $\eta_{KL}={\rm diag}(\pm 1,\dots,\pm 1)$.

   The equations (\ref{f-sum}),(\ref{ea}),(\ref{ma}),(\ref{ds2}),(\ref{fi})
 describe the solution of the equations of motion under
 the following restrictions (\ref{H-cond})--(\ref{myR})
 for functions the $H_R$, incidence matrices and other
 parameters of solution.

 Let us introduce the following designations:
 \begin{equation}
 \label{bel}
   \Box=g^{KL}\partial_K\partial_L,
 \end{equation}
 under the used gauge  conditions
 (Fock-De Donder gauge) $\Box$
 is Beltrami-Laplace operator.
 \begin{equation}
  (\partial f,\partial g)=g^{KL}\partial_K f \partial_L g,
 \end{equation}
 \begin{equation}
  I(R,R')=\left\{
           \begin{array}{lr}
           \sum_{K=0}^{D-1}\Delta_{RK}\Delta_{R'K}, &
               \varsigma_R=-1~~~\mbox{or}~~~\varsigma_{R'}=-1\\
           \sum_{K=0}^{D-1}\Delta_{RK}\Delta_{R'K}-2, &
               \varsigma_R=+1~~~\mbox{and}~~~\varsigma_{R'}=+1\\
           \end{array}
         \right.
 \end{equation}
 for $I(R,R')$ we shall use the term
 {\it intersection index}.

  Using these designations we can write the restrictions for
 the solution parameters in the following form\footnote{
         The important restriction (\ref{pop}) was absent
          in the initial version of this paper
          due to the mistake in calculations.
          This restriction was initially
          introduced by H.~L\"u and C.~N.~Pope
          in the paper \protect\cite{LuPope}.}
 \begin{eqnarray}
 \label{H-cond}
   &&\Box H_R=0,\\
 \label{depen}
   &&\Upsilon_{RK}\partial_K H_R=0,\\
 \label{pop}
   &&\Upsilon_{R'N}\partial_N H_R\Upsilon_{RM}\partial_M H_{R'}=0,\\
 \label{sR}
   &&s_I|_{R~\mbox{\scriptsize describes}~F^{(I)}}
     \prod_{K}(\eta_{KK})^{\Delta_{RK}}=\varsigma_R,\\
 \label{R1}
   &&\ln H_R=\frac{\varsigma_R}{2}
             \sum_{R'}\varsigma_{R'}h_{R'}^2
             \left\{
               I(R,R')-\frac{d_R d_{R'}}{D-2}
               +\frac{\vec\alpha_R\vec\alpha_{R'}}{2}
             \right\}
             \ln H_{R'},\\
 \end{eqnarray}
 \begin{eqnarray}
 \label{myR}
         &&\mbox{if}\;\; I(R,R')=d_R-1=d_{R'}-1,\\
	 && \nonumber
         \mbox{where the indices  $R$ and $R'$ describe the same field
         $F^{(I)}$,}\\
   &&\nonumber \mbox{then}\\
	 &&\nonumber \left\{
		 \begin{array}{l}
         \varsigma_R=\varsigma_{R'}\\
         \sum_{M:\Upsilon_{RM}=\Upsilon_{R'M}=0}
         \partial_M H_R\partial^M H_{R'}=0\\
				 (\partial_M H_R\cdot
          \partial_N H_{R'})
         |_{\Upsilon_{RM}=\Upsilon_{R'N}=0,
            \Upsilon_{RN}=\Upsilon_{R'M}=1}=0
			\end{array}
			\right. ,\\
	 &&\nonumber\mbox{or}\\
	 &&\nonumber\left\{
		 \begin{array}{l}
         \varsigma_R\not=\varsigma_{R'}\\
         (\partial_{M_1} H_R
         \partial_{M_2} H_{R'}-
         \partial_{M_2} H_R
         \partial_{M_2} H_{R'})
         |_{\Upsilon_{R M_1}=\Upsilon_{R M_2}=
            \Upsilon_{R'M_1}=\Upsilon_{R'M_2}=0}=0
			\end{array}
			\right. .
 \end{eqnarray}
 The equation (\ref{sR}) for the standard $(-,+,\dots,+)$
 signature gives restrictions
 $\Delta_{a0}^{(I)}=1$, $\Lambda_{b0}^{(I)}=0$.

 The harmonicity conditions (\ref{H-cond}) will be a subject
 of special discussion in the section \ref{s-bel}.

 The equation (\ref{R1}) in the case of ``independent''
 choice of functions the $H_R$ gives equations for $h_R$
 and the characteristic equations, we shall refer
 this equation as the {\it precharacteristic equation}.

 The energy-momentum tensor can be written in the following
 form
 \begin{equation}
   T_{KL}=\sum_{RR'}T_{KL}^{RR'},
 \end{equation}
 where the terms $T_{KL}^{RR'}$ corresponds to pair of
 branes. The restriction (\ref{myR}) guarantee
 the vanishing of terms $T_{KL}^{RR'}$
 with $R\not=R'$.

\section{Harmonicity conditions}\label{s-bel}

 Equation (\ref{H-cond}) now is not simple condition
 of harmonicity, using equation (\ref{bel}) one can
 rewrite it in the following form
 \begin{equation}
   g^{KL}\partial_K\partial_L H_R=0,
 \end{equation}
 and using (\ref{ds2})
 \begin{equation}
    \sum_{K,L=0}^{D-1}
       \left(
         \prod_{R'\not=R}
           H_R^{h_R^2\Upsilon_{RK}}
       \right)
    \eta^{KL}\partial_K\partial_L H_R=0.
 \end{equation}
 These equations are nonlinear.
 The simplest way to satisfy this equation is
 to set restriction (\ref{plain}) for coordinate
 dependence of the functions $H$.
 In this case
 \begin{equation}
   \eta^{KL}\partial_K\partial_L H_R=0,
 \end{equation}
 i.e. we have several linear differential equations.

 The other simple way to construct $H_R$
 functions is to set
 \begin{equation}
 \label{proizv}
   H_R=\prod_{u}H_{Ru}(x_{a^u_1},\dots,x_{a^u_{n_u}}),~~~~~
   g_{a^u_i a^u_i}=g_{a^u_j a^u_j},~~~~~
   \eta_{MN}\partial_M\partial_N H_{Ru}=0,
 \end{equation}
 taking into account the restriction (\ref{pop}).
 $H_{Ru}$ satisfy restriction (\ref{plain})
 and depend upon different coordinates.
 One can
 find the examples of such solution for 11D supergravity
 in (\ref{ex0-1})--(\ref{r3}) and (\ref{ex11d})--(\ref{ex11L}).

\section{Precharacteristic equations}

 \subsection{Case of independent choice of $H_R$}

 If we want to select the $H_R$ functions
 by an independent way, then, using the precharacteristic
 equations (\ref{R1}), we have
 \begin{eqnarray}
 \label{h-eq}
   \frac{h_R^2}{2}\left\{ d_R-
             \frac{d_R^2}{D-2}+
             \frac{\vec\alpha_R^2}{2} \right\}&=&1,\\
 \label{ch-eq}
   I(R,R')-\frac{d_R d_{R'}}{D-2}+
      \frac{\vec\alpha_R\vec\alpha_{R'}}{2}
      &=&0,~~~R\not=R'.
 \end{eqnarray}
 The equation (\ref{h-eq}) gives the values of $h_R$ in terms
 of the action parameters. The equation (\ref{ch-eq}) is
 the {\it characteristic equation} in the case of independent
 choice of $H_R$. It gives the intersection index.

 \subsection{Case of dependent choice of $H_R$}

 \subsubsection{General statements}

 In this section we assume restriction (\ref{plain})

 In general case one has linear system (\ref{R1})
 of precharacteristic equations for $\ln H_R$
 and conditions $\Box H_R=0$.
 One can construct maximal set of linearly independent
 $-\ln H_R$ functions. Let us numerate elements of this
 set by indices $\varpi,\varrho,\upsilon$, as
 ${\cal C}_{\varpi}$, ${\cal C}_{\varrho}$, etc.

 For arbitrary $C_R=-\ln H_R$ one has
 \begin{equation}
  C_R=\sum_{\varrho=1}^{\Rho} a_{R\varrho}{\cal C}_{\varrho}.
 \end{equation}
 In these terms the equation $\Box H_R=0$ has
 the following form
 \begin{equation}
   \sum_{\varrho,\varpi=1}^{\Rho}a_{R\varrho}a_{R\varpi}
     (\partial {\cal C}_{\varrho},\partial {\cal C}_{\varpi})=
     \sum_{\varrho=1}^{\Rho}a_{R\varrho}\Box {\cal C}_{\varrho}.
 \end{equation}
 Taking into account equations $\Box e^{-\cal C_{\varrho}}=0$
 one can rewrite this equation
 \begin{equation}
   \label{R2}
   \sum_{\varrho\not=\varpi}a_{R\varrho}a_{R\varpi}
     (\partial {\cal C}_{\varrho},\partial {\cal C}_{\varpi})=
     \sum_{\varrho=1}^{\Rho}a_{R\varrho}
     (1-a_{R\varrho})\Box {\cal C}_{\varrho}.
 \end{equation}
 Obviously if $\Rho=1$ the only way to satisfy
 this condition is to set $a_{R\varrho}\in\{0,1\}$.

 \subsubsection{Possible way for solutions construction}

 Similar to above one can find solutions with
 $a_{R\varrho}\in\{0,1\}$ for arbitrary $\Rho$.
 In this case
 $$
   \sum_{\varrho\not=\varpi}
   (\partial {\cal C}_{\varrho},\partial {\cal C}_{\varpi})=0.
 $$
 The simplest way to satisfy this equation is
 to set $(\partial {\cal C}_{\varrho},\partial {\cal C}_{\varpi})=0$
 for every pair $\varrho,\varpi$ for which there exists $R$
 $a_{R\varrho}=a_{R\varpi}=1$. For building of this construction
 using some of $C_R$ functions as ${\cal C}_{\varrho}$
 functions is not necessary, one can use any functions for
 which $\Box e^{-\cal C_{\varrho}}=0$.

 If $s=0$, we can combine only two $\cal{C}_\varrho$ functions.
 In this case one can build solutions using us $\cal{C}_\varrho$
 $a\Re f$ and $b\Im f$, where $a,b=constant$ and $f$ is
 complex analytical function.

 \subsubsection{Examples of solutions families\protect\label{dep}}

 Let us build, for example, the family of such solutions.
 We will discuss the system with one $d+1$-form
 ($N=1$, $d_I=d$) and one or zero dilaton
 ($n=0,1$, $\alpha^{(I)}=\alpha$, if $n=0$
 we will set $\alpha=0$). Let
 \begin{eqnarray}
   l&=&\frac{d^2}{D-2}-\frac{\alpha^2}{2},\\
   I(R,R')&=&I\not=l,~~~~~R\not=R'.
 \end{eqnarray}

 We will consider the case with one electric
 and $\Cal{N}$ magnetic terms in ansatz for
 $d$-form (case with one magnetic and $\Cal{N}$
 electric terms may be constructed totally similarly).
 The subscript $\varepsilon$ designates the electric term,
 and the subscript $\mu$ numerates the magnetic terms.
 Let
 \begin{eqnarray}
   \label{sc}
   C_\varepsilon&=&\sum_{\mu}C_\mu,\\
   (\partial C_{\mu_1},\partial C_{\mu_2})&=&0,
                           ~~~~~\mu_1\not=\mu_2,
 \end{eqnarray}
 where the functions $C_\mu$ are chosen by an
 independent way.
 The equations (\ref{R1}) for the functions $C_\mu$ have the form
 \begin{equation}
   C_\mu=\frac{h_\mu^2}{2}\{d-l\}C_\mu
         +\sum_{\mu'\not=\mu}
         \frac{h_{\mu'}^2}{2}\{I-l\}C_{\mu'}
         -\frac{h_\varepsilon^2}{2}\{I-l\}C_\varepsilon.
 \end{equation}
 Taking into account (\ref{sc}) and independent choice
 of the functions $C_\mu$ we can find
 \begin{equation}
   h_\mu^2=h_\varepsilon^2=h^2=\frac{2}{d-I}.
 \end{equation}
 Now equation (\ref{R1}) for the function $C_\varepsilon$
 \begin{equation}
   C_\varepsilon=\frac{h_\mu^2}{2}\{d-l\}C_\varepsilon
         -\sum_{\mu'}\frac{h_{\mu'}^2}{2}\{I-l\}C_{\mu'}
 \end{equation}
 is the identity.

 We can find another similar solutions family, if
 we consider $\Cal{N}+1$ electric or magnetic terms.
 We will designate one of them by
 the subscript $\varepsilon$, and numerate
 all others by the subscript $\mu$. We
 assume $C_\varepsilon=\sum_{\mu}C_\mu$
 (that is similar to above), and
 \begin{eqnarray}
   I(\varepsilon,\mu)&=&l-c_\varepsilon,\\
   I(\mu_1,\mu_2)&=&l+c_\mu,~~~~~~\mu_1\not=\mu_2.
 \end{eqnarray}

 Similarly to above from the equation (\ref{R1}) for
 the functions $C_\mu$ we can find
 \begin{eqnarray*}
   h_\mu^2 c_\mu&=&h_\varepsilon^2 c_\varepsilon,\\
   \frac{h_\mu^2}{2}\{d-l-c_\mu\}&=&1.
 \end{eqnarray*}
 From the equation (\ref{R1}) for the function $C_\varepsilon$
 we finally find
 \begin{eqnarray}
   c_\mu&=&c_\varepsilon=c,\\
   h_\varepsilon^2&=&h_\mu^2=h^2=\frac{2}{d-l-c}.
 \end{eqnarray}
%%______________________________

 The examples of solutions from these solutions
 families in are (\ref{ex1}), (\ref{ex2}).
 All other nontrivial configurations
 of these families in considered case
 are prohibited by the restriction (\ref{myR}).

\section{The case of 11D supergravity}
 Let us consider in the bosonic sector of
 the 11D supergravity
 \begin{equation}
   I=\frac{1}{2\kappa^2}\int d^{11}x \sqrt{-g}
     \left(R-\frac{F^2}{48}\right)
     +\frac{b}{2\kappa^2}\int A\land F\land F,
 \end{equation}
 The bosonic sector of $D=11$ supergravity consists
 of a metric $g$ and a three-form potential $A$,
 $F=dA$.

 For antisymmetric field and  metric we have the
 following
 \begin{eqnarray}
   F&=&\sum_{a=1}^{E}dA_a+\sum_{b=1}^{M} F_b,\\
   A_{a~M_1 M_2 M_3}&=&
     \epsilon_{M_1 M_2 M_3} h_a  H_a^{-1}
     \prod_{i=1}^{3}\Delta_{aM_i},\\
   F_b^{M_0 M_1 M_2 M_3}&=&
     \epsilon^{M_0 M_1 M_2 M_3 K}
     h_b |g|^{-1/2}
     \partial_K  H_b^{-1}
     \prod_{i=0}^{3}\Lambda_{bM_i}^{(I)},\\
   ds^2&=&\sum_{K,L=0}^{D-1}
           \left(
             \prod_{R}
               H_R^{\varsigma_R h_R^2
                    \left\{
                       \Delta_{RK}-\frac{1}{3}
                    \right\}
                   }
           \right)
            \eta_{KL}dX^K dX^L,
 \end{eqnarray}

 Precharacteristic equation has the following form
 \begin{equation}
 \ln H_R=\frac{\varsigma_R}{2}\sum_{R'}\varsigma_{R'}h_{R'}^2
    \left\{
      I(R,R')-\frac{1}{3}
    \right\}\ln H_{R'}.
 \end{equation}

 If we set $h_R=1$ and $I(R,R')=1,~~R\not=R'$, then
 the precharacteristic equation is an identity.

 For example,
 let us consider the following incidence matrix
$$
 \begin{array}{|c||c|c|c|c|c|c|c|c|c|c|}
        \hline
        0&1&2&3&4&5&6&7&8&9&10\\
        \hline\hline
        \circ&\circ&\circ&&&&&&&&\\
        \hline
        \circ&&&\circ&\circ&&&&&&\\
        \hline
        \circ&&&&&\circ&\circ&&&&\\
        \hline
 \end{array}
$$

 In this incidence table each column represent value of
 world index and each line represent value of
 $R$ index. There is a sign in every box,
 which correspond to $\Delta_{RL}=1$. For electric
 term the sign is ``$\circ$'', and for magnetic
 terms the sign is ``$\bullet$''. A blank box
 corresponds to $\Delta_{RL}=0$.

 The correspondent metric has the form
 \begin{eqnarray}
 \label{ex11d}
   ds^2&=&(H_1H_2H_3)^\frac{1}{3}
          \left(
           -(H_1H_2H_3)^{-1}dx_0^2 \right.\\ \nonumber
          &+&H_1^{-1}(dx_1^2+dx_2^2)
           +H_2^{-1}(dx_3^2+dx_4^2)
           +H_3^{-1}(dx_5^2+dx_6^2)\\ \nonumber
          &+&\left.(dx_7^2+dx_8^2+dx_9^2+dx_{10}^2)  \right).
 \end{eqnarray}
 According to (\ref{proizv}) we can set
 \begin{eqnarray}
 H_1=H_{12}(x_3,x_4)H_{13}(x_5,x_6)H_{10}(x_7,x_8,x_9,x_{10}),\\
 H_2=H_{21}(x_1,x_2)H_{23}(x_5,x_6)H_{20}(x_7,x_8,x_9,x_{10}),\\
 H_3=H_{31}(x_1,x_2)H_{32}(x_3,x_4)H_{30}(x_7,x_8,x_9,x_{10}),\\
 \forall i\not=j\in\{1,2,3\}~:~H_{ij}=1~~\mbox{or}~~H_{ji}=1,\\
 \label{ex11L}
 \partial_K\partial_K H_{Ru}=0,~~~R=1,2,3,~~~u=0,1,2,3.
 \end{eqnarray}

 Let us demonstrate two new solutions
 from the solutions families described
 in previous section.

 These are solutions with $b=0$:
 \begin{itemize}
  \item the case with one electric and two
     magnetic terms, $I=0$, $h^2=2/3$.
$$
 \begin{array}{|c||c|c|c|c|c|c|c|c|c|c|}
        \hline
        0&1&2&3&4&5&6&7&8&9&10\\
        \hline\hline
        \circ&\circ&\circ&&&&&&&&\\
        \hline\hline
        &&&\bullet&\bullet&\bullet&&&&\bullet&\bullet\\
        \hline
        &&&&&&\bullet&\bullet&\bullet&\bullet&\bullet\\
        \hline
 \end{array}
$$
\begin{eqnarray}
\label{ex1}
  ds^2=H_1^{2/9}(H_2H_3)^{4/9}
       \left(
       (H_1H_2H_3)^{-2/3}(-dx_0^2+dx_1^2+dx_2^2)
       \right.\\
\nonumber
      +H_3^{-2/3}(dx_3^2+dx_4^2+dx_5^2)
      +H_2^{-2/3}(dx_6^2+dx_7^2+dx_8^2)\\
\nonumber
      +\left.(dx_9^2+dx_{10}^2)\right)
\end{eqnarray}

  \item the case with three magnetic terms,
     $c=1$, $h^2=2$.
$$
 \begin{array}{|c||c|c|c|c|c|c|c|c|c|c|}
        \hline
        0&1&2&3&4&5&6&7&8&9&10\\
        \hline\hline
        &&\bullet&\bullet&\bullet&&&&&\bullet&\bullet\\
        \hline
        &&&&&\bullet&\bullet&\bullet&&\bullet&\bullet\\
        \hline
        &&&&&&\bullet&\bullet&\bullet&\bullet&\bullet\\
        \hline
 \end{array}
$$
\begin{eqnarray}
\label{ex2}
  ds^2=(H_1H_2H_3)^{4/3}
       \left(
       (H_1H_2H_3)^{-2}(-dx_0^2+dx_1^2)
       \right.\\
\nonumber
      +(H_2H_3)^{-2}(dx_2^2+dx_3^2+dx_4^2)\\
\nonumber
      +(H_1H_3)^{-2}dx_5^2
      +H_1^{-2}(dx_6^2+dx_7^2)
      +(H_1H_2)^{-2}dx_8^2\\
\nonumber
      +\left.(dx_9^2+dx_{10}^2)\right)
\end{eqnarray}

     The last solution exists also for the signatures
     $(+,\dots,+)$ and $(-,-,+,\dots,+)$.
 \end{itemize}

%%------------------------------
 The analysis of supersymmetry conditions is totally similar
to the section ``Supersymmetry in 11D Supergravity'' in
the paper \cite{AIR}, if the brane configuration satisfy the
following condition
\begin{equation}
   I(R,R')=1,~~~R\not=R',
\end{equation}
then one can choose the signs of constants $h_R$ by
a supersymmetric way.

\section{The case of IIB 10D supergravity}

 Let us consider in the bosonic sector of IIB
 supergravity only one field: 5-form $F_5$ (i.e $d=4$),
 then one can write the following action
 \begin{equation}
   I=\frac{1}{2\kappa^2}\int d^{10}x \sqrt{-g}
     \left(R-\frac{1}{2\cdot 5!}F_5^2\right),
 \end{equation}
 with the additional restriction
 \begin{equation}\label{s-d}
   \ast F_5=F_5.
 \end{equation}
 In the p-brane case the equation (\ref{s-d})
 may be rewritten as statement, that there is
 numeration of branes, for which
 \begin{eqnarray}
 \label{IIB-dep}
   && H_a=H_{\overline a},\\
   && h_a=h_{\overline a},\\
 \label{IIB-L-D}
   && \Delta_{aK}+\Lambda_{\overline a K}=1,
 \end{eqnarray}
 where we introduce the pairs of indices $a,\overline a$,
 which correspond to the self-dual pairs of
 the electric and magnetic branes.

 For IIB supergravity theory
 restrictions (\ref{IIB-dep})--(\ref{IIB-L-D})
 are not compatible with the characteristic equations
 (\ref{ch-eq}) for the independent harmonic functions.
 We can write, using the restrictions
 (\ref{IIB-dep})--(\ref{IIB-L-D}),
 the precharacteristic equations (\ref{R1})
 in the following form
 \begin{equation}
   \ln H_a=\sum_{a'}\frac{h_{a'}^2}{2}
           \left\{
             2I(a,a')-d
           \right\}
           \ln H_{a'}.
 \end{equation}
 If the functions $H_a$ are chosen by an independent
 way, then we can find
 \begin{eqnarray}
 \label{hIIB}
   h_a=h_{\overline a}&=&\pm 1/\sqrt{2},\\
 \label{I-IIB}
   I(a,a')&=&2.
 \end{eqnarray}

 Let us consider supersymmetry of such solutions.
 The supersymmetry takes place if there exist
 the Killing spinor $\varepsilon$
 \begin{eqnarray}
   &&\tilde D_M\varepsilon=0,\\
   &&\tilde D_M=\partial_M
               +\frac{1}{4}\omega_{M}^{~~AB}\Gamma_{AB}
               +\frac{i}{8\sqrt{2}\cdot 120}
                \Gamma^{K_1\cdots K_5}\Gamma_M
                F_{K_1\cdots K_5}.
 \end{eqnarray}
 The symbols $\Gamma_A$ are 10D Dirac matrices
 \begin{eqnarray}
   \{\Gamma_A,\Gamma_B\}=2\eta_{AB},\\
   \Gamma_{AB\cdots C}=\Gamma_{[A}\Gamma_{B\cdots}\Gamma_{C]}.
 \end{eqnarray}
 In terms of the considered solutions one can write
 \begin{eqnarray}
   \tilde D_M&=&\partial_M
             -\sum_a \frac{h_a}{4\sqrt{2}}\partial_M\ln H_a
              i(\Gamma(a)+\Gamma(\overline a))\\
   \nonumber
             &-&\sum_a \frac{h_a}{2\sqrt{2}}
              \{2\Delta_{aM}-1\}
              \Gamma_M^{~~N}
              \partial_N\ln H_a
              \frac{
                h_a\sqrt{2}
               +\frac{i}{2}(\Gamma(a)+\Gamma(\overline a))
               }{2},
 \end{eqnarray}
 where
 \begin{eqnarray}
    \Gamma(a)&=&\frac{1}{4!}\prod_{\{A|\Delta_{aA}=1\}}\Gamma_A,\\
    \Gamma(\overline a)&=&
                \frac{1}{6!}\prod_{\{A|\Lambda_{\overline a A}=1\}}
                \Gamma_A.
 \end{eqnarray}
 Using sentence (\ref{hIIB}) we can write
 \begin{equation}
 \label{proIIB}
    \frac{
      h_a\sqrt{2}
     +\frac{i}{2}(\Gamma(a)+\Gamma(\overline a))
    }{2}=
    \pm
    \frac{
      1\pm\frac{i}{2}(\Gamma(a)+\Gamma(\overline a))
    }{2},
 \end{equation}
 i.e. this term is proportional to the sum of two projectors.
 Let us consider the spinor subspace for which
 \begin{equation}
    \Gamma_0\cdots\Gamma_9\cdot\varepsilon=-\varepsilon.
 \end{equation}
 The term (\ref{proIIB}) is a projector for this subspace, so
 similar to 11D case (see \cite{AIR})
 if the brane configuration satisfy the condition \mbox{$I(a,a')=2$,}
 then one can choose the signs of
 constants $h_a$ by a supersymmetric way.

\section*{Acknowledgments}
 The author is grateful to I.~V.~Volovich
 for the statement of problem and for discussion,
 to I.~Ya.~Aref'eva for discussion and
 to H.~L\"u and C.~N.~Pope for indication of
 the mistake in the initial version of this paper.
 This work is partially supported by the RFFI grant
 number 96-01-00312.

%%%%%%%%%%%%%%%%%%%%%
\end{document}